\documentclass[conference,10pt]{IEEEtran}
\usepackage{epsfig,epsf,rotating,setspace,latexsym,amsmath,amssymb,amsfonts,bm,theorem,subfigure,epstopdf}
\usepackage{cite,authblk}
\usepackage{bbm}
\usepackage{algorithm}
\usepackage[noend]{algpseudocode}
\usepackage{color}
\usepackage{mathtools}
\usepackage{soul}

\algrenewcommand\algorithmicforall{\textbf{foreach}}
\algrenewcommand\algorithmicindent{.8em}

\newtheorem{lemma}{Lemma}

\newtheorem{proposition}{Proposition}
\newenvironment{Proof}[1]{\medskip\par\noindent{\bf Proof:\,}\,#1}{{\mbox{\,$\blacksquare$}\par}}

\IEEEoverridecommandlockouts
\allowdisplaybreaks

\begin{document}

\title{Choosing Outdated Information to Achieve \\ Reliability in Age-Based Gossiping}
 
\author{Priyanka Kaswan \qquad Sennur Ulukus\\
        \normalsize Department of Electrical and Computer Engineering\\
        \normalsize University of Maryland, College Park, MD 20742\\
        \normalsize  \emph{pkaswan@umd.edu} \qquad \emph{ulukus@umd.edu}}
        
\maketitle

\begin{abstract}
    We consider a system model with two sources, a reliable source and an unreliable source, who are responsible for disseminating updates regarding a process to an age-based gossip network of $n$ nodes. Nodes wish to have fresh information, however, they have preference for packets that originated at the reliable source and are willing to sacrifice their version age of information by up to $G$ versions to switch from an unreliable packet to a reliable packet. We study how this protocol impacts the prevalence of unreliable packets at nodes in the network and their version age. Using a stochastic hybrid system (SHS) framework, we formulate analytical equations to characterize two quantities: expected fraction of nodes with unreliable packets and expected version age of information at network nodes. We show that as $G$ increases, fewer nodes have unreliable packet, however, their version age increases as well, thereby inducing a freshness-reliability trade-off in the network. We present numerical results to support our findings.
\end{abstract}

\section{Introduction}

\begin{figure}[t]
\centerline{\includegraphics[width=0.8\linewidth]{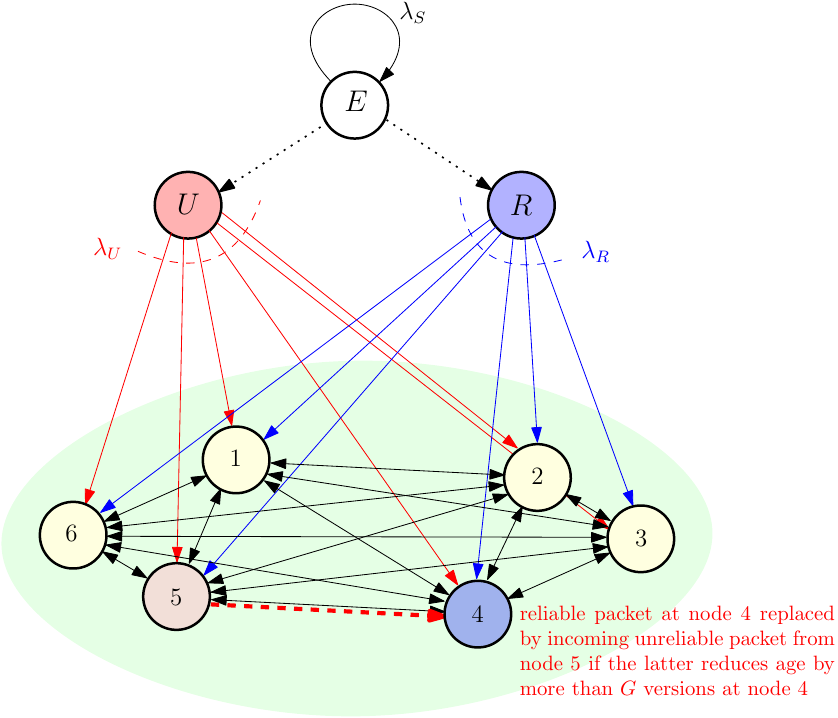}}
\caption{System with reliable ($R$) and unreliable ($U$) sources observing an event ($E$) and gossip network of $6$ nodes, depicting an inter-node transition from node $5$ (possessing unreliable packet) to node $4$ (possessing reliable packet). Node $4$ accepts the incoming packet only if this action results in a decrease in version age at node $4$ by atleast $G$ versions.}
\label{fig:system_model}
\vspace*{-0.4cm}
\end{figure}

We consider a system with a continually updating process or event ($E$) that a network of $n$ nodes, $\mathcal{N}=\{1,\ldots,n\}$, wish to track in real-time, see Fig.~\ref{fig:system_model}. Two sources are available for providing updates about the event to the network, such that information received from one source is considered more reliable than the other. The less reliable source could, for example, be a proxy for a cheap sensor that transmits quantized or noisy measurements to an IoT network. We call the former source as the reliable ($R$) source and the latter source as the unreliable ($U$) source, and the information received from them as reliable and unreliable information, respectively. 

The status of the event is tied to version numbers, such that each time the event gets updated, the version number corresponding to the current state of the event increments by one. The nodes wish to have access to latest version of reliable information about the event. The freshness of information in this work is quantified by version age of information metric \cite{Yates21gossip, kaswan_nonpoisson_version}. Given $V_E(t)$ as the version number corresponding to the current state of event and $V_i(t)$ as the version number of the information about the event present at node $i$, the instantaneous version age of information at node $i$ is defined as $X_i(t)=V_E(t)-V_i(t)$, with $V_E(t)$, and consequently $X_i(t)$, incrementing by one every time the event gets updated.

For faster dissemination of fresh reliable packets, the network employs inter-node gossiping, where each network node regularly transmits the packet in its possession to a neighboring node, chosen uniformly at random each time from the set of neighbors. Gossip protocols have been shown to yield faster dissemination of time-sensitive information in large networks compared to the source single-handedly disseminating timely updates to the network, and have been an active research area in communications and networking \cite{Demers1987EpidemicAF-short, Minsky02cornellthesis, vocking2000, deb2006AlgebraicGossip, Sanghavi2007GossipFileSplit, Yates21gossip, baturalp21comm_struc, kaswan22slicingcoding, mitra_allerton2022,bastopcu_agent_gossip}. 

In Fig.~\ref{fig:system_model}, both reliable and unreliable sources always transmit the latest information about the event. Hence, from the network perspective, they always have the packets corresponding to latest version numbers, and consequently their respective version ages $X_R(t)$ and $X_U(t)$ are zero at all times. User nodes prefer a reliable packet over an unreliable packet even if the former is little bit outdated compared with the latter, and are willing to sacrifice their freshness of information by a small amount if that allows the node to switch to a reliable packet from an unreliable packet. Let $S_i(t)$ represent the reliability status of the packet present at node $i$ at time $t$, with $S_i(t)=0$ implying presence of a reliable packet and $S_i(t)=1$ implying presence of an unreliable packet. When node $i$ sends an update packet to node $j$ at time $t$, node $j$ decides whether to accept or reject the packet based on the following set of rules:
\begin{itemize}
    \item If $S_i(t)=1$ and $S_j(t)=1$, i.e., both nodes possess unreliable information, then node $j$ chooses the packet with lower version age of information.
    \item If $S_i(t)=0$ and $S_j(t)=0$, i.e., both nodes possess reliable information, then node $j$ again chooses the packet with lower version age of information.
    \item If $S_i(t)=0$ and $S_j(t)=1$, i.e., incoming packet has reliable information but node $j$ has unreliable information, then node $j$ will choose the reliable incoming packet as long as $X_i \leq X_j+G$, in other words, the incoming packet is no more than $G$ versions older than the packet already present at node $j$.
    \item If $S_i(t)=1$ and $S_j(t)=0$, i.e., node $j$ already has reliable information packet and the incoming packet is unreliable, then node $j$ would continue to keep its reliable packet as long as $X_j \leq X_i+G$, in other words, the reliable packet present at node $j$ is no more than $G$ versions older than the incoming unreliable packet.
\end{itemize}
With this protocol, let $F(t)$ denote the fraction of user nodes that have unreliable information packet at time $t$, such that 
\begin{align}\label{eqn:F(t)_def}
    F(t)= \frac{S_1(t)+ S_2(t)+ \ldots + S_n(t)}{n}
\end{align}
We are interested in characterizing the long-term expectations $F= \lim_{t \to \infty} \mathbb{E}[F(t)]$ and $x_i= \lim_{t \to \infty} \mathbb{E}[X_i(t)]$, $i \in \mathcal{N}$.

In \cite{kaswan23reliable01}, this protocol was studied for the special case of $G=1$ and $G=0$, where it was seen that the former resulted in higher version age and higher prevalence of unreliable packets compared to the latter. Solving for these special cases in \cite{kaswan23reliable01} involved defining certain test functions in SHS modelling, corresponding either to the reliability status of a set of nodes, or the version age of a set of nodes or certain indicator functions, resulting in recursive equations in one variable to solve for $F$ and $x_i$. In this work, we perform the analysis for general $G$, where we encounter test functions that are products of the above test functions, and the resulting equations are often a recurrence relation in two variables. Since our work deals with the reliability of information, other related gossiping based works are \cite{kaswan23mutation, bastopcu_agent_gossip} regarding information mutation and spread of incorrect information, and \cite{kaswan22timestomping} regarding timestamp manipulation and circulation of outdated packets. 

In the next section, we begin with modelling the problem as an SHS system to derive certain linear equations to characterize $F$ and $x_i$. We then prove several results that allow us to show that $F$ is a decreasing function of $G$ and $x_i$ are an increasing function of $G$. Therefore, $G$ induces a trade-off between reliability and freshness of information. We finally present numerical results to verify our theoretical results.

\section{System Model and SHS Characterization}\label{sec:system_mod}

The system model includes a reliable source ($R$) and an unreliable source ($U$) that send updates to a set of $n$ user nodes $\mathcal{N}={1,\ldots,n}$ about an event ($E$), which gets updated as a rate $\lambda_E$ Poisson process, as shown in Fig.~\ref{fig:system_model}. Both reliable and unreliable sources always have the latest information and send updates to user node $i \in \mathcal{N}$ as Poisson processes with rates of $\frac{\lambda_R}{n}$ and $\frac{\lambda_U}{n}$, respectively. Additionally, for every $i,j\in \mathcal{N}$, node $i$ sends updates to node $j$ as a Poisson process with rate $\lambda_{ij}=\frac{\lambda}{n-1}$ as part of the underlying gossip network. Given the network symmetry, the reliability status and version age processes at all user nodes will be statistically identical, and consequently $x_1=\ldots=x_n$ and $F=s_1$, where $s_i= \lim_{t \to \infty} \mathbb{E}[S_i(t)]$. Next, we use SHS modelling \cite{hespanhashs} to obtain a set of linear equations to derive $s_1$ and $x_1$.

As in \cite{kaswan23reliable01}, for our SHS model, we choose the continuous state as $(\pmb{S}(t),\pmb{X}(t))\in \mathbb{R}^{2n}$, where $\pmb{S}(t)=[S_1(t),\ldots,S_n(t)]$ and $\pmb{X}(t)=[X_1(t),\ldots,X_n(t)]$ represent the instantaneous reliability status and instantaneous version age, respectively, of the $n$ user nodes at time $t$. A transition $(i,j)$ is said to occur when node $i$ sends an update packet to node $j$, with $(E,E)$ denoting an event update, and $(U,i)$ and $(R,i)$ denoting updates to user node $i$ from the unreliable and reliable source, respectively. $S_i(t)$ and $X_i(t)$ remain unchanged between transitions, resulting in the SHS operating in a single discrete mode with the differential equation $(\pmb{\dot S}(t),\pmb{\dot X}(t))=\pmb{0}_{2n}$. The set of transitions is,
\begin{align}\label{eqn:L-set_of_transitions}
    \mathcal{L}= \{&(E,E)\} \bigcup \{(U,i):i \in \mathcal{N}\} \bigcup \{(R,i):i \in \mathcal{N}\}\nonumber \\
    &\bigcup \{(i,j):i,j \in \mathcal{N}\}
\end{align}
such that the transition $(i,j)$ resets the state $(\pmb{S},\pmb{X})$  at time $t$ to $\phi_{i,j}(\pmb{S},\pmb{X},t)\in \mathbb{R}^{2n}$ post transition. The rates $\lambda_{ij}$ associated with each transition $(i,j)$ are given as,
\begin{align} \label{eqn:rates_lambda}
\lambda_{ij} = \begin{cases} 
\frac{\lambda_U}{n}, & i=U,j\in \mathcal{N}\\
\frac{\lambda_R}{n}, & i=R,j\in \mathcal{N}\\
\frac{\lambda}{n-1}, & i,j\in \mathcal{N}\\
\lambda_E, & i=E,j=E
\end{cases}
\end{align}

Next, for a set of nodes $A$, let $R(A)$ and $U(A)$ denote the largest subset of $A$ with reliable information and unreliable information, respectively. Considering a continuous state $(\pmb{S},\pmb{X})$ and a set of nodes $A$, we define the version age of set $A$, denoted as $X_A$, in the following manner: 

\begin{itemize}
    \item If $A=\varnothing$, then $X_A=\infty$.
    \item If $A=R(A)$ or $A=U(A)$, then $X_A=\min_{j \in A}X_j$.   
    \item If $\min_{j \in R(A)}X_j \leq \min_{j \in U(A)}X_j + G$, then $X_A= \min_{j \in R(A)}X_j$.  
    \item If $\min_{j \in U(A)}X_j \leq \min_{j \in R(A)}X_j -G -1 $, then $X_A= \min_{j \in U(A)}X_j$. 
\end{itemize}
 Next, we define reliability status of set $A$, denoted by $S_A$, in the following manner:
\begin{itemize}
    \item If $X_{R(A)} \leq X_{U(A)} + G$, then $S_A=0$.
    \item If $X_{U(A)} \leq X_{R(A)} - G-1 $, then $S_A=1$.
\end{itemize}

In essence, determining $S_A$ and $X_A$ requires us to identify the best node in set $A$ in some sense, such that reliability status and version age of that node are also the reliability status and version age of the set. From the definitions of $S_A$ and $X_A$, we can see that as long as the most recent reliable packet is at most $G$ versions older than the latest unreliable packet in the set of nodes, the node with the latest reliable packet establishes the values of $X_A$ and $S_A$. Otherwise, the node with the latest unreliable packet determines $X_A$ and $S_A$. 

With this definition, considering the transition $(i,j)$ at time $t$, the reset map to $\phi_{i,j}(\pmb{S},\pmb{X},t)=[S_1',\ldots,S_n',X_1',\ldots,X_n']\in \mathbb{R}^{2n}$ can be described as,
\begin{align}
S_{\ell}' &= \begin{cases} 
S_{\{U,\ell\}}, & i=U,j\in \mathcal{N},\ell=j\\
0, & i=R,j\in \mathcal{N},\ell=j\\
S_{\{i,\ell\}}, & i,j\in \mathcal{N},\ell=j\\
S_{\ell}, & \text{otherwise}
\end{cases} \label{eqn:continuous_state_Sl} \\
X_{\ell}' & = \begin{cases} 
X_{\ell}+1, & i=E,j=E,\ell=j\\
\mathbbm{1}_{\{ X_{R(\{\ell\})}=1 \} }, & i=U,j\in \mathcal{N},\ell=j\\
0, & i=R,j\in \mathcal{N},\ell=j\\
X_{\{i,\ell\}}, & i,j\in \mathcal{N},\ell=j\\
X_{\ell}, & \text{otherwise}
\end{cases} \label{eqn:continuous_state_Xl}
\end{align}
where $\mathbbm{1}_{\{.\}}$ represents the indicator function. 

Next, consider a time-invariant test function $\psi:\mathbb{R}^{2n} \to \mathbb{R}$ whose long-term expected value $\mathbb{E}[\psi]=\lim_{t \to \infty} \mathbb{E}[\psi(\pmb{S}(t),\pmb{X}(t))]$ is of interest to us. As in \cite{kaswan23reliable01}, defining $\mathbb{E}[\psi(\phi_{i,j})]=\lim_{t \to \infty} \mathbb{E}[\psi(\phi_{i,j}(\pmb{S}(t),\pmb{X}(t),t))]$, we obtain from \cite[Thm.~1]{hespanhashs}, 
\begin{align} \label{eqn:hespanha_eqn}
    0=\sum_{(i,j)\in \mathcal{L}}\left(\mathbb{E}[\psi(\phi_{i,j})]- \mathbb{E}[\psi] \right)\lambda_{ij}
\end{align}
We will be using this equation repeatedly by introducing a set of time-invariant test functions suitable for our analysis. For more details, readers are encouraged to refer to \cite{hespanhashs,yates_realtime_multisrc,Yates21gossip}.

\section{Reliability and Version Age Analysis}\label{sec:gap_g}

Since version age and reliability status processes are statistically identical for all user nodes, let $A_k$ denote an arbitrary subset of $k$ user nodes. Our first test function is $\psi(\pmb{S},\pmb{X})= S_{A_k}$, which upon $(i,j)$ transition becomes  $\psi(\phi_{i,j}(\pmb{S},\pmb{X},t))=S_{A_k}'$ and can be characterized using (\ref{eqn:continuous_state_Sl}), (\ref{eqn:continuous_state_Xl}) as follows,
\begin{align} \label{eqn:testfunc_resetmap_SAk}
S_{A_k}' = \begin{cases} 
S_{A_k\cup\{U\}}, & i=U,j\in A_k\\
0, & i=R,j\in A_k\\
S_{A_{k+1}}, & i=\mathcal{N} \backslash A_k,j\in A_k\\
S_{A_k}, & \text{otherwise}
\end{cases}
\end{align}
Defining $a_k= \lim_{t \to \infty} \mathbb{E}[S_{A_k}(t)]$ and $b_k= \lim_{t \to \infty} \mathbb{E}[S_{A_k\cup\{U\}}(t)]$, and using (\ref{eqn:hespanha_eqn}) gives,
\begin{align} \label{eqn:hesp_eqn_a_k}
    0=&(b_k- a_k)\frac{k\lambda_U}{n} +(0-a_k) \frac{k\lambda_R}{n} \nonumber\\ 
    &+ (a_{k+1} - a_k)\frac{k(n-k)\lambda}{n-1}
\end{align}
Our second test function is $\psi(\pmb{S},\pmb{X})= S_{A_k\cup\{U\}}$, which has $(i,j)$ transition map as follows,
\begin{align} \label{eqn:testfunc_resetmap_SAkU}
S_{A_k\cup\{U\}}' = \begin{cases} 
1-\mathbbm{1}_{\{ X_{R(A_k)} \leq G-1 \} } , & i=E,j=E\\
0, & i=R,j\in A_k\\
S_{A_{k+1}\cup\{U\}}, & i=\mathcal{N} \backslash A_k,j\in A_k\\
S_{A_k\cup\{U\}}, & \text{otherwise}
\end{cases}
\end{align}
Note that the version age of the most recent unreliable packet in the set $A_k\bigcup\{U\}$ will always remain zero, since the unreliable source has zero version age at all times. Hence, post $(E,E)$ transition, $S_{A_k\cup\{U\}}'$ will be zero, i.e., the node with latest reliable packet will be the best node, only if it had at most $G-1$ version age before transition, i.e., $\mathbbm{1}_{\{ X_{R(A_k)} \leq G-1 \} }$, since version age increments by one after the $(E,E)$ transition.

\begin{figure}[t]
 	\begin{center}
 	\subfigure[]{\includegraphics[width=0.45\linewidth]{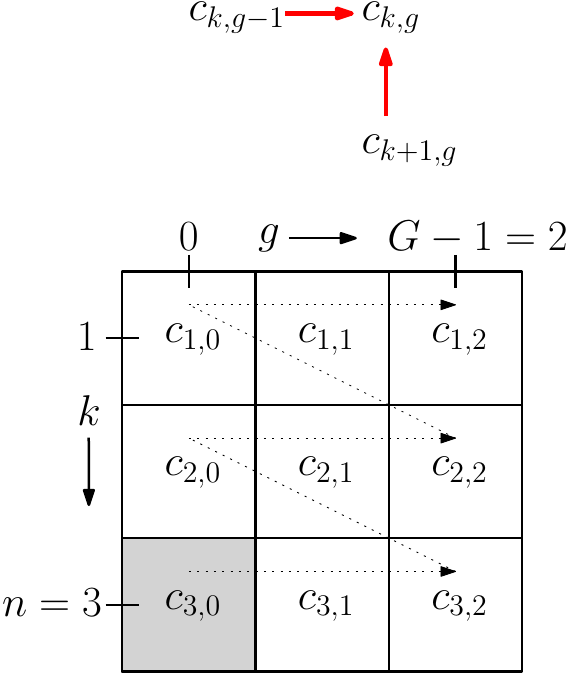}}
         \subfigure[]{\includegraphics[width=0.53\linewidth]{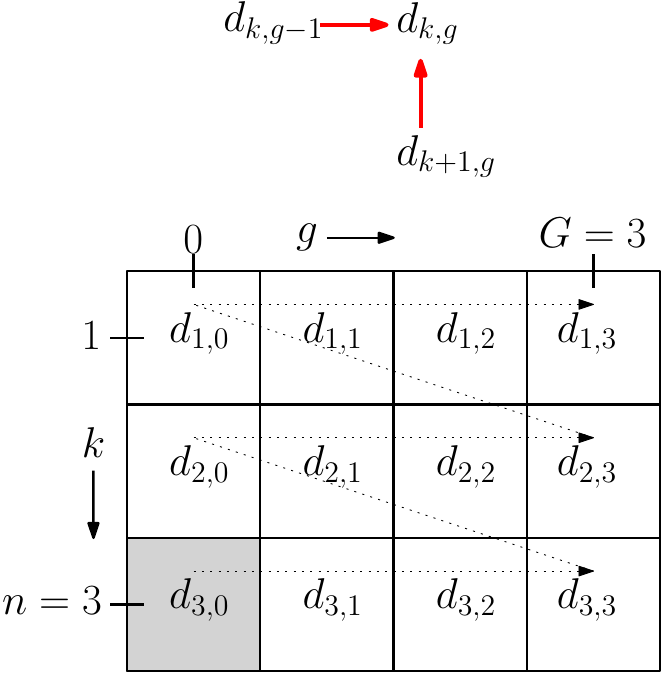}}
 	\end{center}
 	\vspace{-0.3cm}
 	\caption{Approach to computing (a) $c_{k,g}$ (b) $d_{k,g}$.}
 	\label{fig:compute_ckg_dkg}
 	\vspace{-0.3cm}
\end{figure}

Defining $c_{k,g}=\lim_{t \to \infty} \mathbb{E}[\mathbbm{1}_{\{ X_{R(A_k)}(t) \leq g \} }]$, (\ref{eqn:hespanha_eqn}) gives,
\begin{align}\label{eqn:hesp_eqn_b_k}
    0=&(1-c_{k,G-1} - b_k)\lambda_E + (0-b_k)\frac{k\lambda_R}{n}  \nonumber\\ 
    &+(b_{k+1}-b_k)\frac{k(n-k)\lambda}{n-1} 
\end{align}

Our third test function is $\psi(\pmb{S},\pmb{X})= \mathbbm{1}_{\{ X_{R(A_k)} \leq g \} }$, $g\in\{0,\ldots,G-1\}$, which has the $(i,j)$ transition map as follows,
\begin{align} \label{eqn:testfunc_resetmap_1(XRAG=0)}
\mathbbm{1}_{\{ X_{R(A_k)} \leq g \} }' = \begin{cases} 
\mathbbm{1}_{\{ X_{R(A_k)} \leq g-1 \} }, & i=E,j=E\\
1, & i=R,j\in A_k\\
\mathbbm{1}_{\{ X_{R(A_{k+1})} \leq g \} }, & i=\mathcal{N} \backslash A_k,j\in A_k\\
\mathbbm{1}_{\{ X_{R(A_k)} \leq g \} }, & \text{otherwise}
\end{cases}
\end{align}
which, upon employing (\ref{eqn:hespanha_eqn}), gives,
\begin{align}\label{eqn:hesp_eqn_c_kg}
    0=&(c_{k,g-1}-c_{k,g})\lambda_E + (1-c_{k,g})\frac{k\lambda_R}{n} \nonumber\\
    &+ (c_{k+1,g}-c_{k,g})\frac{k(n-k)\lambda}{n-1}
\end{align}
Here, note that version age cannot be a negative number, therefore, $\mathbbm{1}_{\{ X_{R(A_k)} \leq -1 \} } =0$ and $c_{k,-1}=0$. 

Then, equations (\ref{eqn:hesp_eqn_a_k}), (\ref{eqn:hesp_eqn_b_k}), (\ref{eqn:hesp_eqn_c_kg}) can be rewritten as follows,
\begin{align}
    a_k&= \frac{b_k\frac{k\lambda_U}{n} 
    + a_{k+1}\frac{k(n-k)\lambda}{n-1}}{\frac{k\lambda_U}{n} + \frac{k\lambda_R}{n} +  \frac{k(n-k)\lambda}{n-1}} \label{eqn:formula_a_kG}\\
    b_k&= \frac{(1-c_{k,G-1} )\lambda_E 
    +b_{k+1}\frac{k(n-k)\lambda}{n-1} }{\lambda_E + \frac{k\lambda_R}{n} + \frac{k(n-k)\lambda}{n-1}  } \label{eqn:formula_b_kG}\\
    c_{k,g}&=\frac{ c_{k,g-1}\lambda_E +  \frac{k\lambda_R}{n} + c_{k+1,g}\frac{k(n-k)\lambda}{n-1} }{ \lambda_E +   \frac{k\lambda_R}{n} +  \frac{k(n-k)\lambda}{n-1} } \label{eqn:formula_c_kg}
\end{align}
Note that $a_1=s_1=F$ and from (\ref{eqn:formula_a_kG}), we see that computation of $a_1$ requires solving all $a_k$ and $b_k$. In turn, $b_k$ from (\ref{eqn:formula_b_kG}) requires computation of $c_{k,G-1}$ for all $k$, which in turn requires computation of $c_{k,g}$ for all $k$ and $g$ from (\ref{eqn:formula_c_kg}). Therefore, we first compute $c_{k,g}$ as in Fig.~\ref{fig:compute_ckg_dkg}(a), since $c_{k,g}$ depends on $c_{k,g-1}$ and $c_{k+1,g}$. Starting with $c_{n,0}=\frac{\lambda_R}{\lambda_E + \lambda_R}$, for each $k$ in the order $\{n,\ldots,1\}$, we compute $c_{k,g}$ for $g$ in order $\{0,\ldots,G-1\}$. Once we have all the $c_{k,G-1}$, which is the last column of Fig.~\ref{fig:compute_ckg_dkg}(a), we substitute them in (\ref{eqn:formula_b_kG}) to solve for $b_k$ in the order $k=n,\ldots,1$. 
Finally, we use the $b_k$ to compute $a_k$ from (\ref{eqn:formula_a_kG}) in the order $k=n,\ldots,1$.

Next, to characterize the version age at the nodes, we pick the test functions $X_{A_k}$ and $X_{A_k}\mathbbm{1}_{\{ X_{R(A_k)} \leq g \} }$, which have the following $(i,j)$ transition maps
\begin{align} \label{eqn:testfunc_resetmap_XAkG}
X_{A_k}' = \begin{cases} 
X_{A_k}+1, & i=E,j=E\\
X_{A_k}\mathbbm{1}_{\{ X_{R(A_k)} \leq G \} }, & i=U,j\in A_k\\
0, & i=R,j\in A_k\\
X_{A_{k+1}}, & i=\mathcal{N} \backslash A_k,j\in A_k\\
X_{A_k}, & \text{otherwise}
\end{cases}
\end{align}
and
\begin{align} \label{eqn:testfunc_resetmap_XAk1XRAk=1G}
&X_{A_k}\mathbbm{1}_{\{ X_{R(A_k)} \leq g \} }' \nonumber\\
&= \begin{cases} 
(X_{A_k}+1)\mathbbm{1}_{\{ X_{R(A_k)} \leq g-1 \} } , & i=E,j=E\\
0, & i=R,j\in A_k\\
X_{A_{k+1}}\mathbbm{1}_{\{ X_{R(A_{k+1})} \leq g \} }, & i=\mathcal{N} \backslash A_k,j\in A_k\\
X_{A_k}\mathbbm{1}_{\{ X_{R(A_k)} \leq g \} }, & \text{otherwise}
\end{cases}
\end{align}
where (\ref{eqn:testfunc_resetmap_XAk1XRAk=1G}) is obtained from the product of (\ref{eqn:testfunc_resetmap_XAkG}) and (\ref{eqn:testfunc_resetmap_1(XRAG=0)}), however, each case of (\ref{eqn:testfunc_resetmap_XAk1XRAk=1G}) has the same set size $k$ in all product terms, which prevents the need to evaluate any further test cases. Defining $e_k=\lim_{t \to \infty} \mathbb{E}[X_{A_k}(t)]$ and $d_{k,g}=\lim_{t \to \infty} \mathbb{E}[X_{A_k}\mathbbm{1}_{\{ X_{R(A_k)} \leq g \} }]$ and using (\ref{eqn:hespanha_eqn}), these transition maps give the following linear equations,
\begin{align}\label{eqn:hesp_eqn_e_k}
    0=&(e_k+1-e_k)\lambda_E+ (d_{k,G}- e_k)\frac{k\lambda_U}{n} + (0-e_k) \frac{k\lambda_R}{n}  \nonumber\\ 
    &+ (q_{k+1} - e_k)\frac{k(n-k)\lambda}{n-1} 
\end{align}
and
\begin{align}\label{eqn:hesp_eqn_d_k}
    0=&(d_{k,g-1} + c_{k,g-1} - d_{k,g})\lambda_E + (0-d_{k,g})\frac{k\lambda_R}{n}    \nonumber\\ 
    &+(d_{k+1,g}-d_{k,g})\frac{k(n-k)\lambda}{n-1}   
\end{align}
which upon rearrangement, along with (\ref{eqn:formula_c_kg}), give the following set of equations,
\begin{align}
    e_k &= \frac{\lambda_E+ d_{k,G}\frac{k\lambda_U}{n} 
    + e_{k+1} \frac{k(n-k)\lambda}{n-1}}{\frac{k\lambda_U}{n} + \frac{k\lambda_R}{n} +  \frac{k(n-k)\lambda}{n-1}} \label{eqn:formula_e_kG}\\
    d_{k,g} &= \frac{(d_{k,g-1}+c_{k,g-1})\lambda_E 
    +d_{k+1,g}\frac{k(n-k)\lambda}{n-1} }{\lambda_E + \frac{k\lambda_R}{n} + \frac{k(n-k)\lambda}{n-1}  } \label{eqn:formula_d_kg}\\
    c_{k,g}&=\frac{ c_{k,g-1}\lambda_E +  \frac{k\lambda_R}{n} + c_{k+1,g}\frac{k(n-k)\lambda}{n-1} }{ \lambda_E +   \frac{k\lambda_R}{n} +  \frac{k(n-k)\lambda}{n-1} } \label{eqn:formula_c_kg2}
\end{align}

Note that $\mathbbm{1}_{\{ X_{R(A_k)} \leq -1 \} }=0$, and consequently, $d_{k,-1}=0$, since version age is non-negative. To solve for $x_1=e_1$, we first compute $c_{k,g}$ as in Fig.~\ref{fig:compute_ckg_dkg}(a), as discussed previously for solution of $F$. Then, using $c_{k,g}$, we compute $d_{k,g}$ from (\ref{eqn:formula_d_kg}) in the row-wise left to right bottom-up manner shown in Fig.~\ref{fig:compute_ckg_dkg}(b), starting with $d_{n,0}=0$. Finally, using $d_{k,G}$, which is the last column of Fig.~\ref{fig:compute_ckg_dkg}(b), we compute all $e_k$ from (\ref{eqn:formula_e_kG}).

Since $F$ depends on $a_k$, $b_k$ and $c_{k,g}$, we next use (\ref{eqn:formula_a_kG}), (\ref{eqn:formula_b_kG}), (\ref{eqn:formula_c_kg}) to prove certain structural results that enable us to prove that $F$ decreases with $G$. Likewise, we prove certain results using (\ref{eqn:formula_e_kG}), (\ref{eqn:formula_d_kg}), (\ref{eqn:formula_c_kg2}) to show that $x_1$ increases with $G$.  

\begin{proposition}\label{prop:ckg_geq_ckg-1}
    For any $k\in\{1,\ldots,n\}$ and $g\in\{1,\ldots,G-1\}$, we have $c_{k,g} > c_{k,g-1}$.
\end{proposition}

\begin{Proof}
    We provide a proof using double induction on variables $k$ and $g$. We first assume the proposition holds for some $k+1$ and prove for $k$. That is, for some $k\in\{1,\ldots,n-1\}$, we assume $c_{k+1,g} > c_{k+1,g-1}$ for all $g\in\{1,\ldots,G-1\}$ and show that it implies $c_{k,g} > c_{k,g-1}$ for all $g\in\{1,\ldots,G-1\}$. Finally, we prove the proposition for base case of $k=n$.
    
    \textit{Assuming for $(k+1)$ and proving for $k$:}
    To show $c_{k,g} > c_{k,g-1}$ for all $g$, we apply a second induction on variable $g$. We assume the proposition holds for $g-1$, i.e., $c_{k,g-1} > c_{k,g-2}$ and show that it holds for $g$ as well, i.e., $c_{k,g} > c_{k,g-1}$, for some $g\in\{2,\ldots,G-1\}$. Replacing $g$ by $g-1$ in (\ref{eqn:formula_c_kg}), 
    \begin{align}\label{eqn:formula_c_kg-1}
        c_{k,g-1}&=\frac{ c_{k,g-2}\lambda_E +  \frac{k\lambda_R}{n} + c_{k+1,g-1}\frac{k(n-k)\lambda}{n-1} }{ \lambda_E +   \frac{k\lambda_R}{n} +  \frac{k(n-k)\lambda}{n-1} }
    \end{align}
    Subtracting (\ref{eqn:formula_c_kg-1}) from (\ref{eqn:formula_c_kg}), we get
    \begin{align}\label{eqn:diff_c_kg_and_c_kg-1}
        c_{k,g}-c_{k,g-1}=& \frac{ (c_{k,g-1}-c_{k,g-2})\lambda_E }{ \lambda_E +   \frac{k\lambda_R}{n} +  \frac{k(n-k)\lambda}{n-1} } \nonumber\\
        &+ \frac{ (c_{k+1,g}-c_{k+1,g-1})\frac{k(n-k)\lambda}{n-1} }{ \lambda_E +   \frac{k\lambda_R}{n} +  \frac{k(n-k)\lambda}{n-1} }
    \end{align}
    In (\ref{eqn:diff_c_kg_and_c_kg-1}), the first term is positive due to the assumption under the second induction for $g-1$, and the second term is positive due to the assumption under first induction for $k+1$.
    
    For the base case of $g=1$, the analogue of (\ref{eqn:diff_c_kg_and_c_kg-1}) is
    \begin{align}\label{eqn:diff_c_k1_and_c_k0}
        c_{k,1}-c_{k,0}=& \frac{ c_{k,0}\lambda_E }{ \lambda_E +   \frac{k\lambda_R}{n} +  \frac{k(n-k)\lambda}{n-1} } \nonumber\\
        &+ \frac{ (c_{k+1,1}-c_{k+1,0})\frac{k(n-k)\lambda}{n-1} }{ \lambda_E +   \frac{k\lambda_R}{n} +  \frac{k(n-k)\lambda}{n-1} }
    \end{align}
    Here, $c_{k,0} \geq \frac{  \frac{k\lambda_R}{n} }{ \lambda_E +   \frac{k\lambda_R}{n} +  \frac{k(n-k)\lambda}{n-1} } >0$ from (\ref{eqn:formula_c_kg}), which implies that the first term of (\ref{eqn:diff_c_k1_and_c_k0}) is positive. Further, $c_{k+1,1}-c_{k+1,0}>0$ due to the assumption under the first induction for $(k+1)$, which implies the second term of (\ref{eqn:diff_c_k1_and_c_k0}) is positive as well. 

    \textit{Base case of $k=n$:} To prove the proposition for base case of $k=n$, i.e., $c_{n,g} > c_{n,g-1}$ for all $g\in\{1,\ldots,G-1\}$, we use induction on $g$. We assume that it holds for $g-1$, i.e., $c_{n,g-1} > c_{n,g-2}$, and holds for $g$ as well, i.e., $c_{n,g} > c_{n,g-1}$, some $g\in\{2,\ldots,G-1\}$. From (\ref{eqn:formula_c_kg}), we get
    \begin{align}
        c_{n,g}&=\frac{ c_{n,g-1}\lambda_E +  \lambda_R  }{ \lambda_E +   \lambda_R } \label{eqn:cng_}\\
        c_{n,g-1}&=\frac{ c_{n,g-2}\lambda_E +  \lambda_R  }{ \lambda_E +   \lambda_R } \label{eqn:cng-1_}
    \end{align}
    Taking the difference of (\ref{eqn:cng_}) and (\ref{eqn:cng-1_}), and using the assumption under induction for $g-1$, we get
    \begin{align}
        c_{n,g}-c_{n,g-1}= \frac{ (c_{n,g-1}-c_{n,g-2})\lambda_E }{ \lambda_E +   \lambda_R } >0
    \end{align}

    For the base case of $g=1$, similarly (\ref{eqn:formula_c_kg}) gives
    \begin{align}
        c_{n,1}&= \frac{ c_{n,0}\lambda_E +  \lambda_R }{ \lambda_E +  \lambda_R} \label{eqn:c_n1} \\
        c_{n,0}&= \frac{  \lambda_R }{ \lambda_E +  \lambda_R}\label{eqn:c_n0} 
    \end{align}
    From (\ref{eqn:c_n1}) and (\ref{eqn:c_n0}), we get
    \begin{align}
        c_{n,1}-c_{n,0}= \frac{ c_{n,0}\lambda_E }{ \lambda_E +  \lambda_R} = \frac{ \lambda_R\lambda_E +  \lambda_R }{ (\lambda_E +  \lambda_R)^2 } >0
    \end{align}
    
    Hence, $c_{k,g} > c_{k,g-1}$ holds for all $k\in\{1,\ldots,n\}$ and $g\in\{1,\ldots,G-1\}$.
\end{Proof}

\begin{proposition}\label{prop:dkg_geq_dkg-1}
    For any $k\in\{1,\ldots,n\}$, $g\in\{1,\ldots,G\}$, we have $d_{k,g} > d_{k,g-1}$.
\end{proposition}

\begin{Proof}
    We prove the proposition using double induction on variables $k$ and $g$, similar to Proposition~\ref{prop:ckg_geq_ckg-1}. In our first induction, we assume that the proposition holds for $(k+1)$, i.e., $d_{k+1,g} > d_{k+1,g-1}$ and show it holds for $k$ as well, i.e., $d_{k,g} > d_{k,g-1}$, for all $g$. To prove for all $g$, we apply a second induction, where we assume that the proposition holds for $g-1$, i.e, $d_{k,g-1} > d_{k,g-2}$, and show that it holds for $g$ as well, i.e., $d_{k,g} > d_{k,g-1}$.

    Substituting $g-1$ for $g$ in (\ref{eqn:formula_d_kg}), we get
    \begin{align}\label{eqn:dkg-1}
        d_{k,g-1} &= \frac{(d_{k,g-2}+c_{k,g-2})\lambda_E 
    +d_{k+1,g-1}\frac{k(n-k)\lambda}{n-1} }{\lambda_E + \frac{k\lambda_R}{n} + \frac{k(n-k)\lambda}{n-1}  } 
    \end{align}
    Subtracting (\ref{eqn:dkg-1}) from (\ref{eqn:formula_d_kg}), we get
    \begin{align}\label{eqn:d_kg-d_kg-1}
        d_{k,g}-d_{k,g-1}= &\frac{(d_{k,g-1}-d_{k,g-2})\lambda_E}{\lambda_E + \frac{k\lambda_R}{n} + \frac{k(n-k)\lambda}{n-1} } \nonumber\\
        &+\frac{(c_{k,g-1}-c_{k,g-2})\lambda_E}{\lambda_E + \frac{k\lambda_R}{n} + \frac{k(n-k)\lambda}{n-1} } \nonumber\\
        &+\frac{(d_{k+1,g}-d_{k+1,g-1})\frac{k(n-k)\lambda}{n-1}}{\lambda_E + \frac{k\lambda_R}{n} + \frac{k(n-k)\lambda}{n-1} }
    \end{align}
    In (\ref{eqn:d_kg-d_kg-1}), the first term is positive due to the assumption under the second induction for $g-1$, the second term is positive due to Proposition~\ref{prop:ckg_geq_ckg-1}, and the third term is positive due to the assumption under the first induction for $(k+1)$. The base cases of $g=0$ and $k=n$ can be likewise proved.
\end{Proof}

\begin{proposition}\label{prop:bkG_decreases_G}
    $b_k$ is a decreasing function of $G$, for all $k\in\{1,\ldots,n\}$.
\end{proposition}

\begin{Proof}
    For $k=n$, we have from (\ref{eqn:formula_b_kG}), that
    \begin{align}
        b_n&= \frac{(1-c_{n,G-1} )\lambda_E }{\lambda_E +\lambda_R  }
    \end{align}
    We know from Proposition~\ref{prop:ckg_geq_ckg-1} that $c_{n,G-1}$ increases with $G$, and therefore, $b_n$ decreases with $G$. Next, we assume that $b_{k+1}$ is a decreasing function of $G$, and inductively argue that $b_{k}$ is also a decreasing function of $G$. The latter can be seen to be true from (\ref{eqn:formula_b_kG}), where $(1-c_{k,G-1})$ is a decreasing function of $G$ from Proposition~\ref{prop:ckg_geq_ckg-1} and $b_{k+1}$ is a decreasing function of $G$ due to assumption under induction for $(k+1)$.
\end{Proof}

\begin{lemma}\label{lemma:a1_decreases_G}
For given $n\in \mathbb{N}$, $\lambda, \lambda_E, \lambda_R, \lambda_U \in \mathbb{R^{++}}$, $a_1$ is a decreasing function of $G$. Further, $\lim_{G \to \infty}a_1 = 0$.
\end{lemma}

\begin{Proof}
    For $k=n$, we have from (\ref{eqn:formula_a_kG}), that
    \begin{align}
        a_n&= \frac{b_n\lambda_U }{\lambda_U + \lambda_R}
    \end{align}
    We know from Proposition~\ref{prop:bkG_decreases_G} that $b_n$ decreases with $G$, and therefore, $a_n$ decreases with $G$. Next, we assume $a_{k+1}$ is a decreasing function of $G$, and inductively argue $a_{k}$ is also a decreasing function of $G$. From (\ref{eqn:formula_a_kG}), we can see that the denominator does not depend on $G$ and the two terms of the numerator are decreasing functions of $G$ due to Proposition~\ref{prop:bkG_decreases_G} and the assumption under the induction for $(k+1)$.
    
    Therefore, $a_1$ is a monotonically decreasing function of $G$. Further, by definition, $a_1 \in [0,1]$, i.e., $a_1$ is bounded. Thus, by monotone convergence theorem, $\lim_{G \to \infty}a_1$ exists. $G\to \infty$ implies that in the file exchange protocol, nodes do not accept unreliable packets and hold on to their last received reliable packets. Therefore, all network nodes have reliable information at all times and heuristically $\lim_{G \to \infty}a_1$ is zero. 
    
    Mathematically, the same can be shown as follows. First, using (\ref{eqn:hesp_eqn_c_kg}), we can inductively show that $\lim_{G \to \infty}c_{k,G-1}=1$ for all $k$, by iterating over $k$ in the order $\{n,\ldots,1\}$ and using $\lim_{G \to \infty}c_{k,G-1}=\lim_{G \to \infty}c_{k,G-2}$ (the limit $\lim_{g\to \infty}c_{k,g}$ exists by boundedness and monotonicity of $c_{k,g}$ from Proposition~\ref{prop:ckg_geq_ckg-1}). Next, it can be inductively shown using (\ref{eqn:formula_b_kG}) that $\lim_{G \to \infty}b_k=0$ for all $k$, by iterating over $k$ in the order $\{n,\ldots,1\}$. Finally, it can be inductively shown using (\ref{eqn:formula_a_kG}) that $\lim_{G \to \infty}a_k=0$ for all $k$, by iterating over $k$ in the order $\{n,\ldots,1\}$, which completes the proof.
\end{Proof}

\begin{lemma}\label{lemma:e1k_increaeses_G}
For given $n\in \mathbb{N}$, $\lambda, \lambda_E, \lambda_R, \lambda_U \in \mathbb{R^{++}}$, $e_1$ is an increasing function of $G$.
\end{lemma}

\begin{Proof}
    Similar to Lemma~\ref{lemma:a1_decreases_G}, for $k=n$, (\ref{eqn:formula_e_kG}) gives
    \begin{align}
        e_n&= \frac{\lambda_E+d_{n,G}\lambda_U }{\lambda_U + \lambda_R}
    \end{align}
    We know from Proposition~\ref{prop:dkg_geq_dkg-1} that $d_{n,G}$ increases with $G$, and therefore, $e_n$ increases with $G$. Next, we assume $e_{k+1}$ is an increasing function of $G$ to inductively argue $e_{k}$ is also an increasing function of $G$. From (\ref{eqn:formula_e_kG}), we can see that the denominator does not depend on $G$ and the last two terms of the numerator are increasing functions of $G$ due to Proposition~\ref{prop:dkg_geq_dkg-1} and the assumption under induction for $(k+1)$. Therefore, $e_1$ increases with $G$.  
\end{Proof}

$G=0$ implies that reliability status becomes important only when the version ages of a reliable packet and an unreliable packet are the same, i.e., it has no impact on version age of user nodes. Hence, for purposes of calculating $e_1$, we can just assume a total source to network update rate of $\lambda_R+\lambda_U$, which modifies (\ref{eqn:formula_e_kG}) into
\begin{align}
    e_k = \frac{\lambda_E
    + e_{k+1} \frac{k(n-k)\lambda}{n-1}}{\frac{k(\lambda_U+\lambda_R)}{n} + \frac{k(n-k)\lambda}{n-1}}
\end{align}
On the other hand, when $G\to \infty$, packets from the unreliable source are rejected by the network, such that (\ref{eqn:formula_e_kG}) turns into
\begin{align}
    e_k = \frac{\lambda_E
    + e_{k+1} \frac{k(n-k)\lambda}{n-1}}{\frac{k\lambda_R}{n} + \frac{k(n-k)\lambda}{n-1}}
\end{align}
Note that network nodes desire fresh reliable packets, or in mathematical terms, low $x_1=e_1$ and low $F=s_1=a_1$. We note from Lemma~\ref{lemma:a1_decreases_G} and Lemma~\ref{lemma:e1k_increaeses_G} that increasing $G$ leads to decrease in $a_1$, which is desirable, but an increase in $e_1$ which is undesirable. Therefore, there is a trade-off between $x_1$ and $F$ induced by $G$.

\section{Numerical Results}

\begin{figure}[t]
 	\begin{center}
 	\subfigure[]{\includegraphics[width=0.85\linewidth] {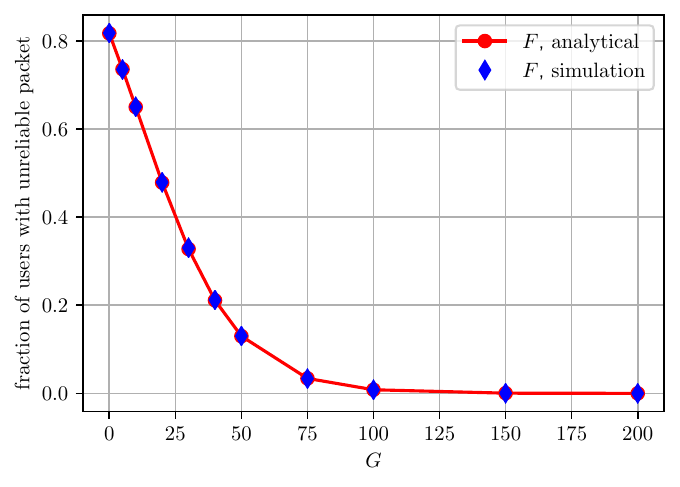}}\\
 	\subfigure[]{\includegraphics[width=0.85\linewidth]{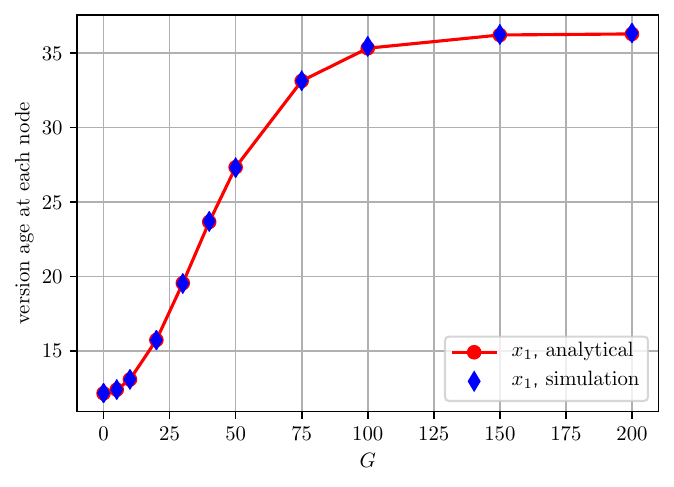}}\\
         \subfigure[]{\includegraphics[width=0.85\linewidth]{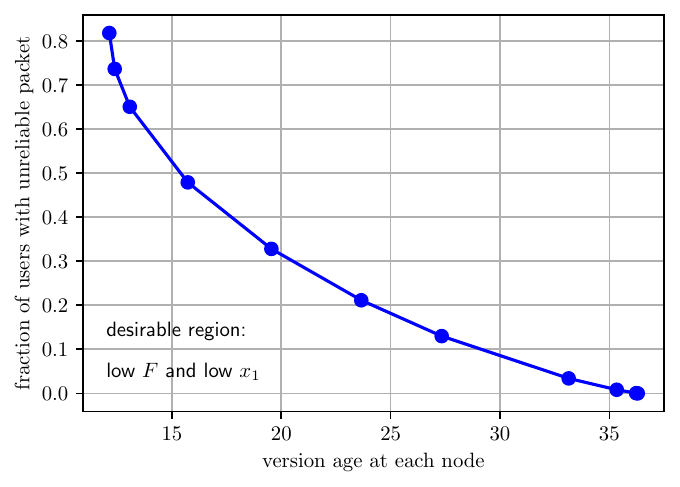}}
 	\end{center}
 	\vspace{-0.1cm}
 	\caption{Analytical and simulation results as a function of $G$ for (a) expected fraction of users with unreliable information $F$, (b) expected version age $x_1$, (c) trade-off between $F$ and $x_1$ induced by $G$.}
 	\label{fig:simulation_vs_formula}
 	\vspace{-0.4cm}
 \end{figure}

We simulate a fully-connected network of $n=50$ nodes with parameters $\lambda_E=2$, $\lambda_U=5$, $\lambda_R=1$ and $\lambda=0.1$ for up to a total time of $10^5$ which we use as proxy for $t \to \infty$. We vary $G$ and plot simulation points (blue dots) of $F$ and $x_1$ on curves (red lines) obtained from equations (\ref{eqn:formula_a_kG}), (\ref{eqn:formula_b_kG}), (\ref{eqn:formula_c_kg}), (\ref{eqn:formula_e_kG}), (\ref{eqn:formula_d_kg}) in Fig.~\ref{fig:simulation_vs_formula}(a) and Fig.~\ref{fig:simulation_vs_formula}(b), respectively. The real-time simulation points coincide with the iterative calculation of the derived equation curves, lending support to the theoretical analysis. Fig.~\ref{fig:simulation_vs_formula}(a) shows that $F$ decreases with $G$ and converges to zero, as suggested by Lemma~\ref{lemma:a1_decreases_G}, and Fig.~\ref{fig:simulation_vs_formula}(b) shows that version age at nodes increases with $G$. Fig.~\ref{fig:simulation_vs_formula}(c) shows the trade-off between $F$ and $x_1$ induced by $G$, with low values of both variables being desirable.

\bibliographystyle{unsrt}
\bibliography{ref_priyanka}

\begin{thebibliography}{10}

\bibitem{Yates21gossip}
R.~D. Yates.
\newblock The age of gossip in networks.
\newblock In {\em IEEE ISIT}, July 2021.

\bibitem{kaswan_nonpoisson_version}
P.~Kaswan and S.~Ulukus.
\newblock Timely tracking of a remote dynamic source via multi-hop renewal updates.
\newblock In {\em IEEE CDC}, December 2024.

\bibitem{Demers1987EpidemicAF-short}
A.~J.~Demers, D.~H.~Greene, C.~H.~Hauser, et~al.
\newblock Epidemic algorithms for replicated database maintenance.
\newblock In {\em ACM PODC}, August 1987.

\bibitem{Minsky02cornellthesis}
Y.~Minsky.
\newblock {\em Spreading Rumors Cheaply, Quickly, and Reliably}.
\newblock PhD thesis, Cornell University, March 2002.

\bibitem{vocking2000}
R.~Karp, C.~Schindelhauer, S.~Shenker, and B.~Vocking.
\newblock Randomized rumor spreading.
\newblock In {\em FOCS}, November 2000.

\bibitem{deb2006AlgebraicGossip}
S.~Deb, M.~Medard, and C.~Choute.
\newblock Algebraic gossip: a network coding approach to optimal multiple rumor mongering.
\newblock {\em IEEE Transactions on Information Theory}, 52(6):2486--2507, June 2006.

\bibitem{Sanghavi2007GossipFileSplit}
S.~Sanghavi, B.~Hajek, and L.~Massoulie.
\newblock Gossiping with multiple messages.
\newblock {\em IEEE Transactions on Information Theory}, 53(12):4640--4654, December 2007.

\bibitem{baturalp21comm_struc}
B.~Buyukates, M.~Bastopcu, and S.~Ulukus.
\newblock Age of gossip in networks with community structure.
\newblock In {\em IEEE SPAWC}, September 2021.

\bibitem{kaswan22slicingcoding}
P.~Kaswan and S.~Ulukus.
\newblock Timely gossiping with file slicing and network coding.
\newblock In {\em IEEE ISIT}, June 2022.

\bibitem{mitra_allerton2022}
P.~Mitra and S.~Ulukus.
\newblock {ASUMAN}: Age sense updating multiple access in networks.
\newblock In {\em Allerton Conference}, September 2022.

\bibitem{bastopcu_agent_gossip}
M.~Bastopcu, S.~R. Etesami, and T.~Bașar.
\newblock The dissemination of time-varying information over networked agents with gossiping.
\newblock In {\em IEEE ISIT}, June 2022.

\bibitem{kaswan23reliable01}
P.~Kaswan and S.~Ulukus.
\newblock Reliable and unreliable sources in age-based gossiping.
\newblock In {\em IEEE ISIT}, June 2023.

\bibitem{kaswan23mutation}
P.~Kaswan and S.~Ulukus.
\newblock Information mutation and spread of misinformation in timely gossip networks.
\newblock In {\em Globecom}, December 2023.

\bibitem{kaswan22timestomping}
P.~Kaswan and S.~Ulukus.
\newblock Susceptibility of age of gossip to timestomping.
\newblock In {\em IEEE ITW}, November 2022.

\bibitem{hespanhashs}
J.~Hespanha.
\newblock Modeling and analysis of stochastic hybrid systems.
\newblock {\em IEE Proc. Control Theory \& Applications, Special Issue on Hybrid Systems}, 153:520--535, January 2007.

\bibitem{yates_realtime_multisrc}
R.~D. Yates and S.~K. Kaul.
\newblock The age of information: Real-time status updating by multiple sources.
\newblock {\em IEEE Transactions on Information Theory}, 65(3):1807--1827, March 2019.

\end{thebibliography}

\end{document}